# A Probabilistic Delay Model for Bidirectional VANETs in City Environments

Md. Mamunur Rashid Akand
Department of Computer Science and Information Technology (CIT)
Islamic University of Technology (IUT)
Board Bazar, Gazipur-1704, Bangladesh
e-mail: mrakand@hotmail.com

Mir Tafseer Nayeem
Department of Computer Science and Information Technology (CIT)
Islamic University of Technology (IUT)
Board Bazar, Gazipur-1704, Bangladesh
e-mail: mtnayeem@yahoo.com

Md. Rokon Uz Zaman Sumon
Department of Computer Science and Information Technology (CIT)
Islamic University of Technology (IUT)
Board Bazar, Gazipur-1704, Bangladesh
e-mail: rokon1120@yahoo.com

Muhammad Mahbub Alam
Department of Computer Science and Information Technology (CIT)
Islamic University of Technology (IUT)
Board Bazar, Gazipur-1704, Bangladesh
e-mail: mma@iut-dhaka.edu

*Abstract*— **Routing in VANETs (Vehicular Ad hoc NETworks) is a challenging task due to large network sizes, rapidly changing topology and frequent network disconnections. State-of-the-art routing protocols tried to address these specific problems especially in city environments (vehicles constrained by road geometry, signal transmissions blocked by obstacles, degree of congestion in roads etc). It was noticed that in city scenarios co-directional roads consist of a collection of disconnected clusters because of traffic control strategies (e.g., RSU (Road Side Units), stop signs and traffic lights). In this paper, we propose an inter-vehicle ad-hoc routing metric called EFD (Expected Forwarding Delay) based on the vehicular traffic statistics (e.g., densities and velocities) collected on-the-fly. We derive an analytical expression for the expected size of a cluster in co-directional traffic. In case of disconnection between two co-directional clusters the opposite directional clusters are used as a bridge to propagate a message in the actual forwarding direction to reduce the delay due to carry and forward. Through theoretical analysis and extensive simulation, it is shown that our link delay model provides the accurate link delay estimation in bidirectional city environments.**

## I. Introduction

Vehicular Ad hoc Networks (VANETs) is a subclass of mobile ad hoc networks (MANETs) with special mobility pattern and rapidly changing topology. So the existing routing protocol of MANETs cannot be directly applied to VANETs. VANET is a representative model for IVC. Inter-vehicle communications (IVC) has been gaining a great deal of importance over the past few years. To support the development the US FCC (Federal Communications Commission) has allocated 75 MHz in the 5.9 GHz band for licensed Dedicated Short Range Communication (DSRC) [2] and IEEE has defined a new standard for DSRC named IEEE 802.11p. In recent years, the radio range of VANETs is extended to almost 1,000 meters. This has encouraged lots of governments and prominent industrial corporations such as Toyota, BMW and Daimler-Chrysler to launch several projects like Advanced Driver Assistance Systems (ADASE2) [3], Crash Avoidance Metrics Partnership (CAMP) [4], CarTALK2000 [5], FleetNet [6], and DEMO 2000 by Japan Automobile Research Institute (JSK).

Along with the recent developments in the VANET field several commercial applications (e.g., hotels, restaurants and parking space availability, announcements of sale information, deliver advertisements, remaining stock at a department store etc) help to reduce the extra time and fuel wasted by the drivers and passengers while traveling, entertainment applications (e.g., Internet access and multimedia content sharing ) have been envisioned. In these types of applications the users (e.g., passengers or drivers) can tolerate up to seconds or minutes of delay as long as the reply will finally return.

## II. Related Works

This section highlights major attempts made in routing protocols in VANET scenarios.

In previous works there have some major attempts in applying conventional MANET routing protocols to VANETs. On-demand approaches such as AODV [9] or DSR [10] suffer from broadcast storm problem.

Position-based routing has proven to be well suited for highly dynamic environment due to the low cost and popularity of global positioning system (GPS) and Geo-Location Services. In geographic routing data are routed to vehicles based on their geographic location. Examples for position-based routing algorithms are face-2 [15] and GPSR [16]. Among them GPSR (which is algorithmically identical to face-2) is seems to be scalable and well suited for very dynamic networks. In GPSR, greedy forwarding is used to send packets to nodes that are always progressively closer to the destination. However, there are some cases where packets will reach a local maximum.

Naumov et al. [12] presented the Advanced Greedy Forwarding (AGF) and also incorporated a velocity vector of speed and direction to accurately determine the location of a destination that significantly improves the effectiveness as well as the performance of GPSR [16]. Naumov et al. [12] also introduced Preferred Group Broadcasting (PGB) with route auto-correction strategy to improve AODV.

To deal with the challenges of city scenarios, Lochert et al. [3] proposed GSR, a position-based routing with topological information. This approach employs greedy forwarding along previously selected shortest path. Simulation results show that GSR outperforms topology based approaches like (AODV and DSR ) with respect to packet delivery ratio and latency by using realistic vehicular traffic. Later Lochert et al. [11] also designed GPCR without the help of map information, which is similar to GSR [3] but does not rely on planarization of nodes. GPCR [11] employs a restricted greedy forwarding strategy which has a better recovery strategy than the perimeter mode of GPSR [16]. However, both of the protocols didn't consider the case of low traffic density and vehicles' movement, which make it difficult to find an end-to-end connection along the pre-selected path thus it failed to maintain route stability.

MDDV [7] and VADD [1] are two multi-hop routing protocols; the idea is without an end-to-end connection the message can be delivered through carry and forward, to the destination. When a network disconnection occurs, nodes carry the packet with itself and forward the packet to the nearest neighbor that moves into its vicinity or communication range. VADD only considers how to find a path from a mobile vehicle to a coffee shops where the destination is static and proposes a delay model which is over simplified. However, when the vehicle density is low, the optimal path may not always be available at the moment. Thus, VADD has to deliver packets via detoured paths. In the worst case, the packet may go through a much longer path that's why VADD experiences dramatic performance degradation in packet delivery delay, and MDDV even renders poor reliability.

## III. PROBLEM FORMULATION

In this section, we formulate the data forwarding in vehicular networks based on the following assumptions.

### A. Assumptions

1) Many commercial navigation service vendors such as Garmin Ltd, MapMechanics [14] and Yahoo Maps provide automatic/periodic updates of traffic conditions such as vehicle density, vehicle arrival rate $\lambda$ and average vehicle speed $v$ per road segment.
2) In case of large populated or urban areas during night hours very low density and high speed traffic ($v_{\max}$). On the other hand rush-hour traffic has low speed ($v_{\min}$) with high volume. For the sake of simplicity our delay model assumes that each vehicle has an independent speed taken uniformly from the interval $[v_{\min}, v_{\max}]$ and travels at this constant speed $v_c$ independently from other vehicles.

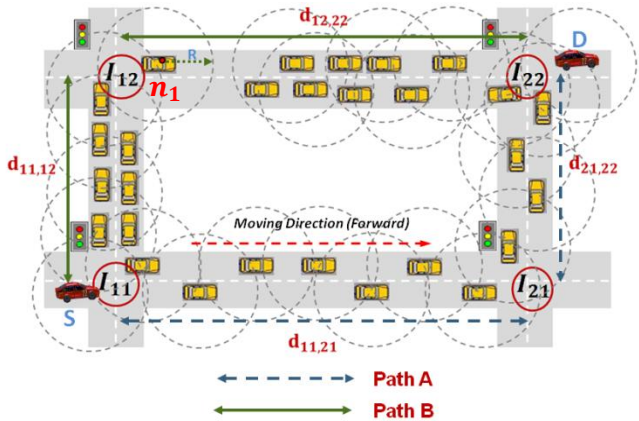

Fig. 1. Packet forwarding scenarios.

Let's consider the packet forwarding scenarios described in fig. 1 where source S wants to communicate with destination denoted by D. There are two alternate paths from source intersection $I_{11}$ such as $I_{11} \to I_{12} \to I_{22}$ or $I_{11} \to I_{21} \to I_{22}$ to reach at $I_{22}$ which is the closest intersection to the destination D. Where two paths have the same distance from $I_{11}$ to $I_{22}$, that means $l_{11,12} + l_{12,22} = l_{11,21} + l_{21,22}$. On the other hand, path B ($I_{11} \to I_{12} \to I_{22}$) has higher network density than path A ($I_{11} \to I_{21} \to I_{22}$).

We know that, Network density = $\frac{\text{number of vehicles}}{\text{road segment length}}$

$$\rho_{11,22}(I_{21}) = \frac{11}{(l_{11,21} + l_{21,22})}$$ For path A if we use $I_{21}$ as an intermediate intersection.

$$\rho_{11,22}(I_{12}) = \frac{17}{(l_{11,12} + l_{12,22})}$$ For path B if we use $I_{12}$ as an intermediate intersection.

Surely, we can see that $\rho_{11,22}(I_{12}) > \rho_{11,22}(I_{21})$ as $l_{11,12} + l_{12,22} = l_{11,21} + l_{21,22}$ but the packet forwarding delay is less in path A. That means, $d_{11,21} + d_{21,22} < d_{11,12} + d_{12,22}$ since path B has the temporary network fragmentation problem .That's why packet carrier $n_1$ in path B needs to carry the packet further to overcome the link breakage .On the other hand, path A has well connectivity hence data packets can be forwarded by multi-hop wireless transmission manner. The carry delay is the dominating part of the total forwarding delay because carry delay is several times longer than the multi-hop communication delay. For example, a vehicle takes 90 seconds to travel along a road segment of 1mile with a speed of 40MPH; however, it takes only 10 milliseconds to forward a packet over the same road segment.

The forwarding delay depends on the inter-vehicle distance which is exponentially distributed with parameter $\lambda$. The authors of [8, 13] found that an exponential model is a good fit for urban vehicular traffic in terms of inter-vehicle distance and time distribution. These two distributions both combinedly define the connectivity of the forwarding path segments.

## IV. EFD: LINK DELAY MODEL

In this section we analyze the link delay for one road segment with bidirectional vehicular traffic with the arrival rate $\lambda$, vehicle speed $v$, road segment length $L$ and the communication range $R$.

In this paper, we define,

1. *Connected Component or Cluster* connected group of vehicles that can communicate with each other via one-hop or multi-hop communication.
2. *Expected Forwarding Delay (EFD) as* the expected time taken by a packet carrier to forward a data packet through VANET to a moving destination vehicle.
3. *Disconnection length($l_d$)* when a packet carrier doesn't find any suitable next hop in its communication range $R$, thus it carry the data packet with itself to overcome the disconnection.
4. *Connection Length($l_c$)* when a data packet is forwarded by multi-hop communication among vehicles through connected component.

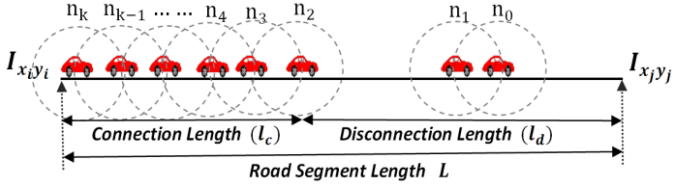

Fig. 2. One way road is used for calculating the forwarding distance.

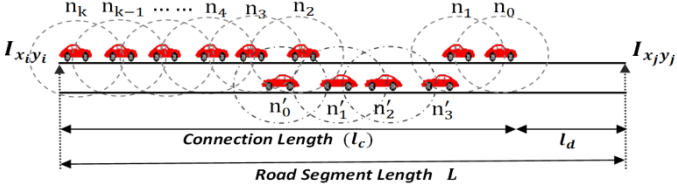

Fig. 3. Bidirectional road is used for calculating the forwarding distance.

In VADD [1] one way road segment is used to calculate the forwarding delay. As shown in fig. 2 disconnection occurs in vehicle $n_2$, therefore vehicle $n_2$ needs to carry the data packet with itself. As the carry delay is significantly larger than the multi-hop delay this will also make the total forwarding delay larger. To reduce the delay further, we have used cluster in the opposite direction as bridges to fill this gap between the clusters in the same direction. The proposed scheme will have less delay than the VADD [1], as we can see in the fig 3 that the disconnection length $l_d$ has significantly reduced compare to in fig. 2. As the carry distance is the dominating part in the total forwarding delay here the carry delay is reduced in fig. 3 by using the opposite directional cluster.

### A. Expected Forwarding Delay in a Cluster

Expected forwarding delay in a cluster is derived in 4 steps as follows.

#### 1. Determining expected number of vehicle in a cluster

A group of vehicles form a cluster if inter-vehicle distance X between any two successive vehicles in that group does not exceed the transmission range R as in fig. 4.

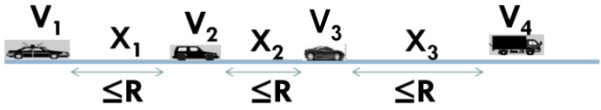

Fig. 4. Inter-vehicle distances in a cluster.

If $\Pr\{X \leq R\}$ defines the probability that inter-vehicle distance X does not exceed the vehicle transmission range R, When X exceeds R, we reach the end of a cluster. If it requires V-1 number of vehicles to reach the end of a cluster, then we can determine the probability that V number of vehicles are inside a cluster using geometric distribution as follows.

$$\Pr(V) = (1 - \Pr\{X \leq R\}) \cdot \Pr\{X \leq R\}^{V-1}, V \geq 1 \quad (1)$$

Now, we can write:-

$$\Pr\{X \leq R\} = 1 - e^{-\lambda r} \quad (2)$$

$$E[X] = \frac{1 - e^{-\lambda r}(\lambda R + 1)}{\lambda(1 - e^{-\lambda r})} \quad (3)$$

Where, E[X] is the expected inter vehicle distance which is a truncated exponential random variable [13].

From equation (1) and (2) we can find out the expected number of vehicle in a cluster-

$$E[V] = \frac{1}{1 - \Pr\{X \leq R\}} = \frac{1}{e^{-\lambda r}} \quad (4)$$

#### 2. Determining expected length of the cluster

Inter-vehicle distance X is independent and identically distributed random variable [8, 13] with truncated exponential distribution. Number of vehicle V is also a random variable. Therefore, we can use Wald's equation to determine expected length of cluster as follows-

$$E[C] = E\left[\sum_{i=1}^{V-1} X_i\right] = E[V - 1] \times E[X] \quad (5)$$

#### 3. Determining Expected Hop Count in a Cluster

Say a vehicle S needs to forward the packet to the next vehicle; if we are lucky in Advanced Greedy Forwarding (AGF) [12] then we will find the next packet carrier sitting exactly in the transmission range(R) boundary of S. If this is the case for every packet carrier in the path then it is the best case and we will find minimum hop count.

So, the minimum number of hop count in a cluster

$$H_{min} = \frac{\text{Expected Cluster Size}}{\text{Transmission Range}} = \frac{E[C]}{R}$$

On the other hand, if we consider every next vehicle as the carrier and forward the packet to them, this will be the worst case. So, the hop count maximum ($H_{max}$) will be,

$$H_{max} = \frac{\text{Expected Cluster Size}}{\text{Expected Inter Vehicle Distance}} = \frac{E[C]}{E[X]}$$

However, we will not find every vehicle in the boundary of transmission range(R) for each packet, forwarding to next hops. In order to make our research more realistic in average case we have taken the average of these $H_{min}$ and $H_{max}$.

Therefore, Expected hop count $E[H] = \frac{H_{max} + H_{min}}{2} \quad (6)$

#### 4. Determining Expected Forwarding Delay in a Cluster

Now we have computed expected hop count E[H] and we know per hop delay $D_h$. From this information, we can determine Expected Forwarding Delay $E[d_c]$ in a cluster-

$$E[d_c] = E[H] \times d_h \quad (7)$$

### B. Delay due to Carry and forward

There can be three cases due to a disconnection in the road segment.

#### 1. Best Case

In Fig. 5 cluster d wants to forward the packet to cluster g but there is a disconnection between cluster d and g as the distance between d and g means $X_{d,g} > R$. As there is an opposite directional cluster f within the transmission range (R) of both d and g, which can relay the data packet from d to g. Here, $Y_1$ is the carry distance, which will be zero in this particular case. The probability of this situation is-

$$P_1 = \Pr\{X_{d,f} \leq R\} \Pr\{X_{f,g} \leq R\}$$
$$Y_1 = 0$$
$$f_{Y_1}(y) = \begin{cases} 1, & y = 0 \\ 0, & \text{otherwise} \end{cases}$$

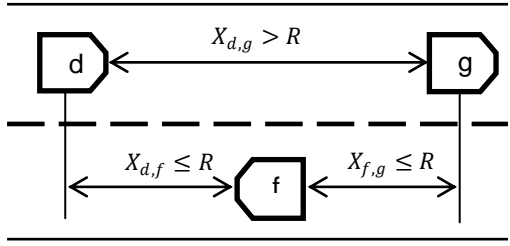

Fig. 5. Best Case in using opposite directional traffic as bridge.

### 2. Average Case

In Fig. 6 there are disconnections between d and g and also in between d and f, but f is connected to g. However, there is a probability that d and f will be connected as they are moving towards one another. Moreover, our main goal is to forward the data packet from d to g via f, we also need to make sure that f and g remain in contact as earlier. So the possible distance cluster f can move is $a = R - X_{f,g}$, as f and g are moving away from each other with constant speed; the carry distance will be $Y_2 = \frac{a}{2}$. The probability of this situation is-

$$P_2 = \Pr\{X_{d,f} > R\} \Pr\{X_{f,g} \le R\}$$
$$a = R - X_{f,g}$$
$$Y_2 = \frac{a}{2}$$
$$f_{Y_2}(y) = \begin{cases} \dfrac{\lambda e^{-\lambda y}}{1 - e^{\lambda(R+2a)}}, & \text{when } y = \dfrac{a}{2} \\ 0, & \text{otherwise} \end{cases}$$

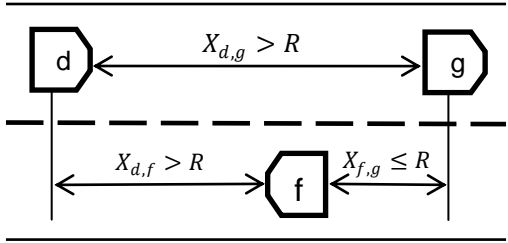

Fig. 6. Worst Case in using opposite directional traffic as bridge.

### 3. Worst Case

In Fig. 7 there are disconnections between d and g, and also in f and g but d and f are in contact. So, cluster d will transmit the data packet to f, as $X_{f,g} > R$ cluster f fails to relay it to g. In this case, cluster f will store the data packet in its buffer and forwards it back to d when they encounter each other. So, the packet carrying distance of cluster f will be $Y_3 = \frac{X_{d,f}}{2}$ as they are moving towards each other with constant speed. The probability of this situation is-

$$P_3 = \Pr\{X_{d,f} \le R\} \Pr\{X_{f,g} > R\}$$
$$Y_3 = \frac{X_{d,f}}{2}$$
$$f_{y_3}(y) = \begin{cases} 1, & \text{when } y = X_{d,f}/2 \\ 0, & \text{otherwise} \end{cases}$$

Based on above 3 cases, the density function of the disconnection length ($l_d$) is as follows, here, i = Case Number.
$$l_d = f_Y(y) = \sum_{i=1}^{3} P_i \times f_{Y_i}(y) \qquad (8)$$

From the Fig. 3. We know that,
$$l_t = l_c + l_d$$

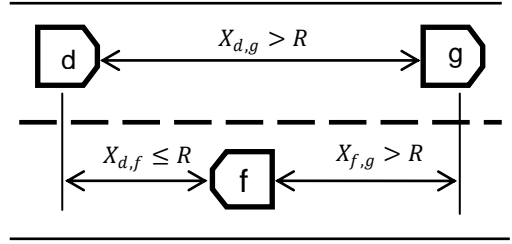

Fig. 7. Worst Case in using opposite directional traffic as bridge.

In realistic scenarios, there can be more than one connection length ($l_c$) and disconnection length ($l_d$) in the total road segment of length $l$. Total delay in a particular road segment will be series of connection ($d(l_c)$) and disconnection delays ($d(l_d)$) up to the road segment of length ($l$). Now, we define the total delay in road segment length $l$ ($d_t(l)$) recursively as follows, here $v$ = velocity:-

$$d_t(l) = d(l_c) + d(l_d) + d_t(l - (l_c + l_d))$$
$$d_t(l) = E[d_c] + \frac{l_d}{v} + d_t(l - (l_c + l_d))$$

Here, the recursive function of $d_t(l)$ will terminate when we will reach $d_t(l \le 0)$. Thus, we have formulated the Expected Forwarding Delay (EFD) in a road segment of length ($l$).

## V. PERFORMANCE EVALUATION

In this section, we evaluate the performance of EFD by comparing it with a state-of-the-art scheme, VADD [1]. The evaluation is based on the following:

1. **Performance Metric:** We use expected forwarding delay as performance metric.
2. **Parameters:** We investigate the impact of vehicular traffic density.

TABLE I. SIMULATION CONFIGURATION

| Parameter | Description |
| --- | --- |
| Road Network | The number of intersections is 25. The area of the road map is 4.2miles×3.7miles. |
| Communication Range | R = 250 meters (i.e., 820 feet). |
| Number of vehicles | The number N of vehicles moving within the road network. The default of N is 100. |
| Time-To-Live | The expiration time of a packet. The default of TTL is ∞ (i.e., no timeout). |

A road network with 25 intersections is used in the simulation. Each vehicle's movement pattern is determined by a random waypoint model. During the simulation, following an exponential distribution with a mean of 4 seconds, packets are dynamically generated from 15 vehicles in the road network. The total number of generated packets is 50,000 and the simulation is continued until all of these packets are either delivered or dropped due to TTL expiration. The default system parameters are used those specified in Table I.

## A. Forwarding Behavior Comparison

We compare the forwarding behaviors of EFD and VADD with the cumulative distribution function (CDF) of the actual packet forwarding delays. From Figure 8, it is very clear that EFD has smaller packet delivery delay than VADD. For any given packet deliver delay, EFD always has a larger CDF value than VADD before they both reach 100% CDF. For example, TBD reaches 90% CDF with a delivery delay of about 580 seconds while the value for VADD is about 800 seconds.

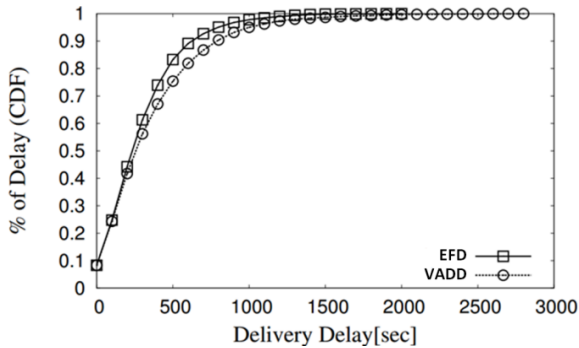

Fig. 8. CDF Comparison for Delivery Delay

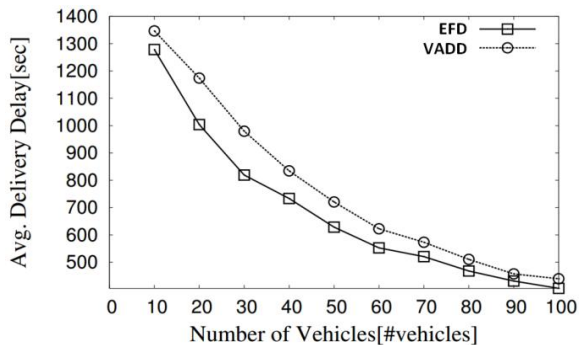

Fig. 9. Impact of the Number of Vehicles

## B. The Impact of Vehicle Number N

The number of vehicles in the road network determines the vehicular traffic density in a road network. Through our extensive simulations, we observe that under low vehicular traffic density, EFD significantly outperforms VADD in terms of packet forwarding delay.

Figure 9 shows the packet forwarding delay comparison between EFD and VADD with varying number of vehicles. As shown in Figure 9, EFD has smaller packet delivery delay than VADD at all vehicular densities. The smallest delay reduction is 5% at N = 10 while the largest delay reduction is 16.5% at N = 30. However, in the sparse road networks (N <10), by using both the bidirectional road and the vehicular traffic statistics, EFD has an average of 10.3% delivery delay reduction (from N = 10 to N = 100) over VADD, which only considers the vehicular traffic statistics.

## VI. CONCLUSION & FUTURE WORKS

In this paper, we propose an inter-vehicle ad-hoc routing metric called EFD (Expected Forwarding Delay). Our future work will focus on the end to end delay estimation between the source and destination in the road network topology graph (RNTG) by introducing city blocks which is a more challenging task.